\documentclass{article}
\usepackage[T1]{fontenc}
\usepackage[utf8]{inputenc}
\usepackage{ismir} 
\usepackage{amsmath,cite,url}
\usepackage{graphicx}
\usepackage{color}
\usepackage{booktabs}
\usepackage{multirow}
\usepackage{nicefrac}
\usepackage{microtype}

\title{Masked diffusion enables coherent beat tracking}

\multauthor{Francesco Foscarin \hspace{1.5cm}Filip Korzeniowski\hspace{1.5cm} Richard Vogl}
{Moises AI\\
{\tt firstname.lastname@moises.ai}}

\def\authorname{F. Foscarin, F. Korzeniowski, and R. Vogl}

\usepackage[bookmarks=false,pdfauthor={\authorname},pdfsubject={\pdfsubject},hidelinks]{hyperref}

\begin{document}

\maketitle

\begin{abstract}
Current neural networks for beat tracking generate invalid outputs, such as consecutive downbeats and erratic tempo changes, even when these are not present in the training data.
Heavy post-processing techniques can alleviate these problems, but the original cause of this inconsistent behaviour remains unknown.
We hypothesise that it stems from inadequate modelling of multiple plausible output beat grids, resulting in an invalid mixture of competing interpretations.
We propose a masked diffusion approach that properly models multiple outputs and enables the model to build coherent predictions through iterative inference.
We devise three modifications to standard masked diffusion that enable its application to beat tracking: independent masking of beats and downbeats during training and inference, a balanced masking scheduler for inference, and peak-picking across inference steps.
Our approach reduces erratic behaviours and improves beat-tracking performance.
\end{abstract}

\section{Introduction}\label{sec:introduction}

\begin{figure}
    \centering
    \includegraphics{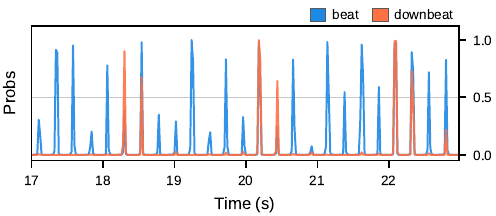}
    \caption{A particularly problematic output of the Beat This system~\cite{foscarin2024beatthis} from the GTZAN rock\_00029. It contains erratic beat tempo doubling/halving and consecutive downbeats.}
    \label{fig:problematic}
\end{figure}

The task of audio beat and downbeat tracking has been deeply explored for decades, and in recent years~\cite{ganghui2026,bolt2026,foscarin2024beatthis,ganghui2025pre,ganghui2025hinge,gagnere2025,zhaobeat,sapinto2023challenging,lostanlen2025stomp,morais2025,fichtinger2025}, and a variety of architectures, losses, data augmentation techniques, pre-training and fine-tuning techniques have been proposed.
All these works use neural networks (NNs) in a \textit{one-step} approach: for each temporal frame, a beat activation is predicted in a single forward pass.

However, this approach implicitly assumes that, given an input, there is exactly one valid output.
When the musical content is ambiguous, the model might produce predictions that mix different solutions and do not correspond to any single coherent interpretation.
Beat tracking suffers from this problem; for some songs, listeners may adopt different metrical levels, and alternative, stylistically plausible beat locations are possible~\cite{mckinney2006ambiguity,lartillot2023hardanger,cano2021sesquialtera}. This ambiguity is also partially accounted for in beat evaluation metrics~\cite{beatles}. 
For such problematic songs, we can clearly observe the aforementioned incoherent behaviour in the beat trackers NN predictions, even those developed specifically to work without a heavy post-processing phase, like the Beat This model~\cite{foscarin2024beatthis} (see Figure~\ref{fig:problematic} for an example). 

Post-processing techniques, such as the Dynamic Bayesian Network (DBN)~\cite{bock2016joint}, fix erratic outputs by enforcing strict and simplistic constraints, such as a fixed number of beats per bar and limited tempo variations. While this alleviates symptoms, it does not address the underlying cause in the NN, which can degrade performance in both non-DBN and DBN systems (e.g., by metrical level switching~\cite{chiu2022analysis}).\footnote{Moreover, the DBN post-processing is inherently unable to handle complex music with a variety of time signatures, time signature changes, and sudden severe tempo variations~\cite{foscarin2024beatthis}.} Since the DBN masks the issue we want to address, we focus on a non-DBN approach in this paper.


The proposed solution to incoherent NN beat predictions is a model that supports multiple valid interpretations and consistently selects one at inference time.
Autoregressive models achieve this by generating outputs step by step, thereby modelling the joint distribution over the full output sequence. However, implementing an autoregressive beat tracker requires a set of architectural choices, each with its own limitations. A frame-wise autoregressive decoder would require one prediction per frame, making inference very slow (e.g., 1500 passes for 30 seconds at 50 fps). A more compact \verb|[beat-type, beat-time]| tokenisation avoids this, but then absolute-time encoding\footnote{This is the approach of the Whisper~\cite{radford2023robust} speech transcription system, and Murgul and Heizmann\cite{murgul2025beat}  MIDI beat tracker.} requires many time tokens and does not work for longer excerpts, while relative-time encoding is prone to error accumulation during generation~\cite{murgul2025beat}.


We take a different approach to let the model consider multiple output variants while avoiding the drawbacks mentioned above: it requires only a few forward passes, maintains the unproblematic frame-wise encoding of the output, and requires only minor modifications to well-tested beat-tracking networks.
We adopt a \textit{Masked Diffusion Model} (MDM), which accepts not only audio as input but also partial output sequences (i.e., beats and downbeats in our case). 
During inference, the full output is built by iteratively taking partial predictions along with the audio as input and producing additional predictions (see Figure~\ref{fig:inference}). This way, the model can select a valid output during the early steps and then coherently fill the rest of the output sequence.

While MDM is widely used with minimal variation across deep learning domains such as language and images, its adaptation for beat tracking poses specific challenges.
1) \emph{the output label imbalance}: the frame-wise annotations mainly consist of non-events with a few sparse beat and downbeat events (the ratio is $\approx90$ to $1$ for downbeats at 50 fps on our training data); 2) the \emph{multi-task problem formulation}: we are simultaneously classifying beats and downbeats; 3) the \emph{lack of temporal precision} for beat/downbeat, which cause a network trained with impulse-like peaks to produce wide peak predictions.

In this paper, we describe the fundamental design choices required to make the MDM approach work for beat tracking and show that MDM drastically reduces erratic behaviour and substantially improves performance, especially on metrics that are heavily influenced by it. 

\section{Related work}\label{sec:related_work}

The backbone of this work is the Beat This model from Foscarin et al.~\cite{foscarin2024beatthis}, whose entire weights, training code, and data are open-source. We use the same data, augmentation methods, metrics, and most hyperparameters, and adapt their model and their shift-tolerant weighted loss to work with masked diffusion.


Gagneré et al.~\cite{gagnere2025} is the only NN-based beat-tracking paper to consider the aforementioned issue of multiple valid outputs in their approach. They use multiple heads in their model, which aim to capture different hierarchical levels. However, this idea is only used during their self-supervised pretraining stage; during fine-tuning and inference, they use a standard single-head architecture for beat and downbeat prediction, so their predictions still suffer from the problems we described in the introduction. We still include this model in our evaluation as one of the high-performing systems.
Two other papers with high performance are by Ru et al.~\cite{ganghui2025pre,ganghui2025hinge}, who present fine-tuning techniques to adapt large pretrained audio models for beat tracking. We consider the latter paper in our comparison as it presents, on average, slightly better results.

From the MDM literature, we draw heavily on LLaDA by Nie et al.~\cite{nie2025llada}, which presents a simple, easy-to-adapt framework that achieves strong results on large models, comparable to those of autoregressive methods. We adapt their training and inference techniques to address the specific challenges of our beat-tracking scenario.\footnote{For readers more familiar with the MaskGIT approach~\cite{chang2022maskgit}, which was also applied to spectrogram generation~\cite{comunita2024specmaskgit}, the two approaches are closely related~\cite{zheng2024masked}. We chose the LLaDA MDM mostly because it has a simpler implementation that uses linear masking during both training and inference.}

\section{Method}\label{sec:method}
Given a sequence $x$, generative models aim to approximate the true but unknown distribution $p_{data}(x)$ by optimising the distribution $p_\theta(x)$ parametrised by a model with parameters $\theta$. MDMs define this distribution through a forward process that gradually masks positions and a reverse process that recovers them \cite{austin2021structured}. 

Since our beat tracking data are not just sequences, but a set of pairs $(x,y)$, where $x$ is the input audio, and $y$ is the beat sequence, we need to reformulate the distribution above as the conditional distribution $p_\theta(y|x)$. During the forward and backward processes, only $y$ is masked and unmasked, while $x$ remains fully visible at all steps.
In the following, we describe our model, training, and inference setup, and compare them with both existing language MDMs and beat trackers.

\subsection{Model}
As mentioned in Section~\ref{sec:related_work}, this work is based on the Beat~This architecture~\cite{foscarin2024beatthis}, which processes a 30-second log-mel-spectrogram input with a frontend block mixing convolutions and directional time-frequency attention, and then passes it through a 6-layer RoFormer. In this section, we describe the modifications required to turn it into a Masked Diffusion Model. One of our objectives is to keep the required architectural changes minimal, enabling us to leverage previous research on architectures, losses, hyperparameter optimisation, and post-processing techniques.

\begin{figure}
  \centering
  \includegraphics[alt={Model image},width=0.9\linewidth]{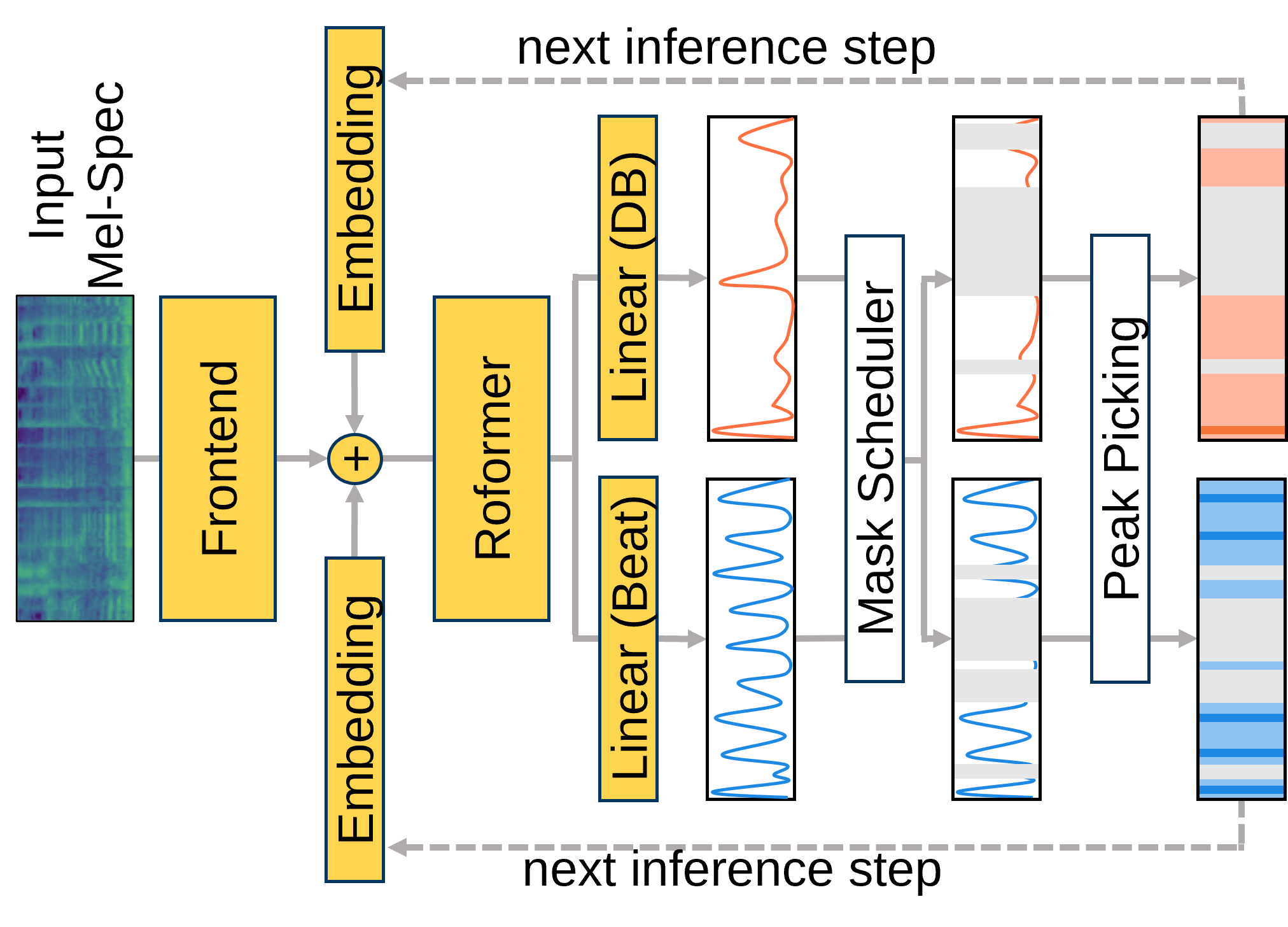}
  \caption{Masked diffusion adaptation of the Beat This model (yellow blocks). We also display the path to reuse previous predictions during subsequent inference steps.}
  \label{fig:model}
\end{figure}

\subsubsection{Token Embeddings}
We add two learnable embedding tables, one for beats and one for downbeats, each mapping a vocabulary of four token types \{\texttt{MASK}, \texttt{NO\_EVENT}, \texttt{EVENT}, \texttt{PAD}\} to the transformer dimension $d$. While it is possible to have a single table for both beats and downbeats by splitting the \texttt{EVENT} token into two \texttt{BEAT} and \texttt{DOWNBEAT} tokens, it would require restructuring our problem from two binary predictions to a multi-class prediction. This would make it difficult to train on datasets that lack downbeat annotations, and require a new loss, since the one from~\cite{foscarin2024beatthis} is only binary.
For these reasons, we keep the two binary beat and downbeat predictions separate.

Unlike language MDMs, our model must handle two inputs from different modalities: audio and beats/downbeats. Assuming that audio requires pre-processing to extract meaningful features, we dedicate the frontend to audio only, whereas beat and downbeat embeddings are added directly before the 6-layer RoFormer (see Figure~\ref{fig:model}). Before adding, we rescale them by $\sqrt{d}$, where $d$ is the dimension of the transformer hidden layer. This increases the importance of the beats input over the audio, and proved simple and effective in preliminary experiments. 

\subsubsection{Output Head}
The original Beat~This model uses a \emph{sum head}: a single linear layer that projects the transformer output into two channels (beat and downbeat logits), with the final beat channel computed as the sum of the two. This helps predict a beat for every downbeat.
However, we observed that this sometimes creates artifacts such as inverted beat spikes in the downbeat logits.
While this does not affect one-step prediction, in our case, the masking scheduler relies on the logit confidences for remasking. 
To avoid potential issues, we simplify the output to two independent linear heads (one per channel) and ensure to predict a beat for every downbeat during the unmasking process instead.

\subsubsection{Activation Function}
We substitute GELU activations with SwiGLU~\cite{swiglu_paper} in the feed-forward network (FFN) (an inverted bottleneck that projects from $d$ to $4d$, computes activations, and projects back to $d$) after each attention block. SwiGLU uses half of its input dimension as a learnable gate. To maintain a comparable parameter count, we reduce the FFN dimension to 2/3, as suggested by the original paper. Best practice (e.g.,~\cite{chowdhery2023palm}) recommends rounding up the hidden dimension to a power of 2 to keep GPU processing efficient (we choose 32 due to our small $d$).
We keep the GELU between convolutions in the frontend.
Using SwiGLU is not necessary for our MDM to work, but it aligns our architectural design with well-tested modern networks and improves our results, at no extra complexity.

\begin{figure*}
    \centering
    \includegraphics{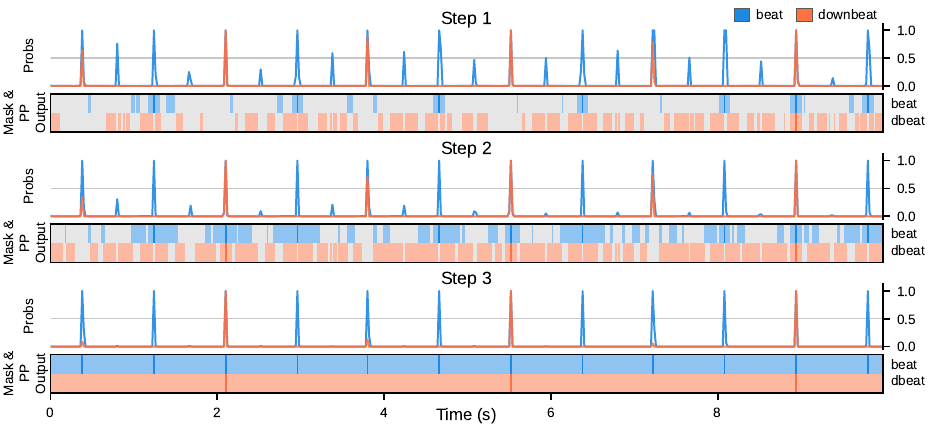}
    \caption{First 10 (of 30) seconds of a 3-step inference on GTZAN pop\_00050. The transformer outputs probabilities for beats and downbeats, and at each step, the masking scheduler unmasks 1/3 of the most confident positive predictions, along with a balanced number of negative predictions (see Section~\ref{sec:balanced}). These are processed by the Peak Picking (PP) module (see Section~\ref{sec:pp}) and fed into the transformer (along with the audio) for the next prediction step.}
    \label{fig:inference}
\end{figure*}

\subsection{Training}\label{sec:training}
We train with the standard masked diffusion objective of MDMs~\cite{nie2025llada}. Given a ground-truth sequence $\textbf{y}$ of length $N$, at each training step we sample a masking probability $m$ uniformly from the interval $[\epsilon, 1]$, where $\epsilon$ is a small number to ensure training stability. Every frame $y_i \in \textbf{y}$ is then independently masked with probability $m$ (i.e., we replace $y_i$ with \texttt{MASK}) and fed into the network along with the audio. The  loss is computed between the original ground truth $y$ and predictions $\hat{y}$ only over the masked positions:
\begin{equation}
  \mathcal{\textbf{L}}_\text{diff}(  \textbf{y},\hat{\textbf{y}}) = \frac{1}{m} \sum_{i \in \text{masked}} \mathcal{L}( y_i,\hat{y}_i),
\label{eq:loss}
\end{equation}

\subsubsection{Masking}
Unlike in language MDMs, we have two distinct output channels associated with separate ground truth (beat and downbeat). Crucially, the two channels are masked \emph{independently}: each draws its own ratio and its own mask, so at any given training step the model may see, e.g., 30\% of beat frames masked and 80\% of downbeat frames masked.
This is a prerequisite for the confidence-based unmasking strategy to work correctly at inference time for each channel (see Section~\ref{sec:inference}).
Additionally, we introduce a hyperparameter $p_\text{onlymask}$ that sets the probability of fully masking both channels, forcing the model to predict the entire sequence from audio alone.

\subsubsection{Loss}
A big advantage of the loss formulation in Equation~\ref{eq:loss} is that we can simply drop in the Shift-tolerant weighted BCE loss~\cite{foscarin2024beatthis} as $\mathcal{L}$ on the right side. This loss was shown to sharpen output logit peaks and balance logit magnitudes, both of which are highly desirable properties during inference. Separated losses are computed for beats and downbeats with equal weight.

We have to deal with an additional source of complexity: the use of padding when processing pieces shorter than 30 seconds. We only mask and compute the loss over unpadded frames, and average it over the number of unpadded elements $N$. This can cause numerical instability when the loss denominators become too small, so we use the actual ratio of masked frames for $m$ (rather than the masking probability), 
and force the minimum number of masked frames (which depends on $\epsilon$ and the total number of unpadded frames) to never be lower than that of a fully unpadded piece.

\subsection{Inference}\label{sec:inference}

Inference starts from a fully masked sequence of $N$ frames and iteratively reveals tokens over $S$ steps.
At each step, the model performs a forward pass conditioned on the currently revealed tokens; the masking scheduler unmasks a subset of the remaining masked frames (i.e., assigns their predicted token values) until the entire sequence is unmasked. While we can find in the literature many ways to select the subset of frames to unmask at every step, we consider a widely used and simple one: given $S$ steps, we unmask at each step the $N/S$ most confident masked predictions~\cite{nie2025llada}. For simplicity, we do not sample from the distribution; instead, we pick the token with the maximum probability.
Adapting this generic masked diffusion loop to beat tracking required three non-obvious design choices.

\subsubsection{Balanced Unmasking Schedule}\label{sec:balanced}
Given one channel output logits, we compute the prediction confidence as $|\text{logits}|$. However, we notice that negative logits (corresponding to non-event frames) are generally more confident than positive ones. Therefore, a pure confidence-based strategy would reveal non-event frames first, leading to a train-inference mismatch, since during training, the model always sees a balanced mix of event and non-event frames. 
Iterative inference in this setting decreases performance rather than improving it. 

To solve this problem, we propose a novel \emph{balanced} unmasking schedule.
At each step, we compute the current positive-to-negative ratio among the model predictions for the masked frames and split the unmasking budget $N/S$ accordingly between predicted positives and negatives.

\subsubsection{Independent Per-Channel Unmasking}
We experimented with different ways of aggregating the confidence scores from the two channels before deciding which positions to unmask, as well as with simple strategies such as unmasking first beats, and then downbeats, or vice versa. Iterative inference in these settings leads to lower performance. 
We hypothesise that the reason is that there are pieces in which beats are simpler to produce than downbeats (e.g., pieces with stable beats and time signature changes), and vice versa (e.g., ethereal music with clear harmonic changes at downbeats). Moreover, beats and downbeats are very different signals: downbeats are approximately 4 times sparser than beats, and, even with positive weights in the loss, their logit magnitudes lie on different scales.

The solution is independent channel unmasking, enabled by our independent masking during training (Section~\ref{sec:training}).
Because each channel was always masked and revealed separately, the model's per-channel confidence scores are individually meaningful, calibrated to each channel's sparsity, and can naturally influence one another in an order determined by the model. 
We therefore run the balanced unmasking schedule independently for beats and downbeats, each with its own confidence scores and negative-to-positive ratio.

\subsubsection{Peak Picking Between Steps}\label{sec:pp}
Frame-wise beat tracking networks tend to predict activations that spread across several adjacent frames near each true beat.
The shift-tolerant weighted BCE loss produces sharper peaks than standard BCE, but clusters of adjacent positive frames still appear at inference time (see, for example, the first beat in Figure~\ref{fig:problematic}). The Beat~This paper uses a peak-picking post-processing step that suppresses all non-maximum positions within a sliding 7-frame proximity window before thresholding to get the final predictions.

In our masked diffusion iterative inference, nearby peaks produce nearby \texttt{EVENT} tokens, a situation that never occurred during training, since in the training data, each event is a single impulse in the non-event sequence.
We address this problem by applying the aforementioned peak-picking post-processing after each inference step.
Whenever a new \texttt{EVENT} is unmasked, all frames within the proximity window are also unmasked and forced to \texttt{NO\_EVENT}, preventing adjacent frames from being revealed as positives in subsequent steps.

We also take the opportunity to enforce musically meaningful behaviour: Every time we unmask a downbeat \texttt{EVENT} at a frame position, we also unmask a beat \texttt{EVENT} at the corresponding position, and force all neighbours to \texttt{NO\_EVENT}. If there is already an unmasked beat \texttt{EVENT} in the proximity window, we move it to the downbeat position (though this situation never happened in our experiments, since the model learns to predict overlapping beats and downbeats). Unlike the sum head of~\cite{foscarin2024beatthis} that only supports this behaviour, this process enforces that every downbeat is also a beat. Note that, unlike the DBN, we enforce a much smaller set of rules that are universally valid across all kinds of music. Figure~\ref{fig:inference} shows an example of our iterative inference.

\subsubsection{Model Ensembling}
NNs are known to be poor at estimating probabilities and tend to be overconfident~\cite{guo2017calibration}. A more reliable measure of confidence can be obtained by averaging the output logits across multiple models. 
We test this approach by averaging the output activations of multiple ensembled models after each inference step, before computing the prediction confidence for unmasking. Model ensembling is a well-tested technique in beat tracking but typically not used in the MDM literature, likely due to the inference cost of large models. For our $\approx$25M parameter model, we find the cost acceptable.

\section{Experiments}

Our experiments aim mainly at isolating the effect of the MDM paradigm, specifically how it affects the coherence of the results. To this end, we focus on CMLt and AMLt metrics, as they measure the stability of inter-beat intervals and are therefore strongly affected by incoherent tempo variations and/or time signatures.

The training data setup is similar to~\cite{foscarin2024beatthis}: first, we optimise hyperparameters on a validation part 
of our training data; then we train on the full training data (incl. the validation set) and use a hold-out test set to report results.
We provide the 8-fold cross-validation metrics across all training datasets, and all our predictions as supplementary material.\footnote{\url{https://github.com/fosfrancesco/md_beat_this}}

\subsection{Experimental Settings}\label{sec:experimental_settings}
We employ the same datasets as the Beat This paper~\cite{foscarin2024beatthis}:
we train and validate on 4556 tracks, including 
those without downbeat annotations, and test on the GTZAN~\cite{gtzan} dataset (993 pieces, excluding one unannotated track and 6 tracks that miss downbeat annotations).

\begin{figure}
    \centering
    \includegraphics{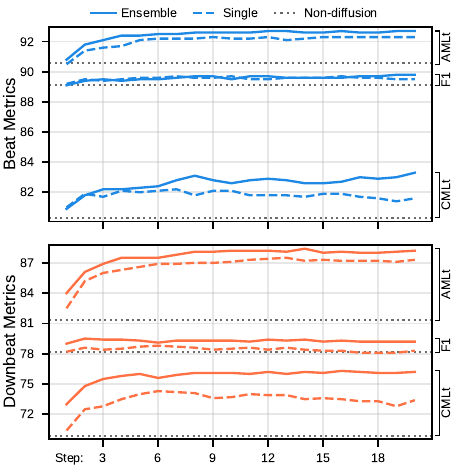}
    \caption{Effect of our MDM formulation, number of inference steps and ensembling on performance.}
    \label{fig:step}
\end{figure} 

\begin{table*}[t]
    \centering
    \begin{tabular}{llllllll}
        \toprule
         & & \multicolumn{3}{c}{Beat} & \multicolumn{3}{c}{Downbeat}\\
         \cmidrule(lr){3-5}
         \cmidrule(lr){6-8}
         & & \multicolumn{1}{c}{F1} & \multicolumn{1}{c}{CMLt} & \multicolumn{1}{c}{AMLt} & \multicolumn{1}{c}{F1} & \multicolumn{1}{c}{CMLt} & \multicolumn{1}{c}{AMLt} \\
         \midrule
         \multirow{3}{*}{\textit{No DBN}}
         & Beat This~\cite{foscarin2024beatthis}  & $89.1 \pm 0.3 $ &  $79.8 \pm 0.6 $ &  $89.8 \pm 0.4 $ &  $78.3 \pm 0.4 $ &  $67.3 \pm 0.8 $ &  $79.1 \pm 0.6 $  \\
         & Gagner\'e ST-BCE~\cite{gagnere2025}  & $89.6$ & $81.8$ & $91.8$ & $77.5$ & $67.0$ & $80.8$ \\
         & \textbf{Ours (8 inference steps)}      & $\textbf{89.7} \pm \textbf{0.2}$ & $\textbf{82.9} \pm \textbf{0.4}$ & $\textbf{92.5} \pm \textbf{0.2}$ & $\textbf{79.5} \pm \textbf{0.3}$ & $\textbf{76.4} \pm \textbf{0.6}$ & $\textbf{88.5} \pm \textbf{0.3}$ \\
         \midrule
         \multirow{2}{*}{\textit{DBN}}
         & Gagner\'e BCE~\cite{gagnere2025}        & $\textbf{89.6}$ & $\textbf{82.6}$ & $92.5$ & $78.3$ & $\textbf{74.7}$ & $88.2$ \\
         & MusicFM+HingeNet~\cite{ganghui2025hinge} & $89.2$ & $80.9$ & $\textbf{93.7}$ & $\textbf{79.8}$ & $73.2$ & $\textbf{89.5}$ \\
         \bottomrule
    \end{tabular}
    \caption{Comparison of SOTA beat tracking systems on GTZAN.
    Results for \cite{foscarin2024beatthis,gagnere2025,ganghui2025hinge} are taken from their papers.}
    \label{tab:soa}
\end{table*}


We train with $p_\text{onlymask}=0.4$, and $\epsilon=0.05$ (see Section~\ref{sec:training}). Language MDMs use a much lower $\epsilon$, e.g., 0.001~\cite{nie2025llada}. However, we train on a shorter sequence (1500 vs 4096) and deal with short padded pieces, which increases the risk of numerical instability.\footnote{This is less problematic when training in bf16 mixed precision, but we train on fp16 mixed due to GPU limitations.} For the same reason, we deviate from \cite{foscarin2024beatthis} and reduce the learning rate to $0.0004$, increase the weight decay to 0.1 (excl.\ embedding layers), introduce gradient clipping of 1, and decrease AdamW $\beta_2$ to 0.95~\cite{smol}.
We also double the number of training epochs from 150 to 300 and employ the WSD learning rate scheduler~\cite{wsd}, with 1000 steps of warm-up and 15\% of the total steps for linear decay.


We compute metrics using the \texttt{mir\_eval} package~\cite{raffel2014mir_eval} with default parameters. Note that, while we enabled our model to disambiguate between valid competing outputs, the metrics still assume that only one solution is correct. AMLt partially address this problem by considering half/double tempi, but the default configurations don't include ternary subdivisions, and complex time signatures, tempo changes, and ambiguous beat positions are not handled.


\subsection{Effect of MDM Formulation}

Figure~\ref{fig:step} compares a single \& ensemble MDMs with a non-diffusion version of the model. The latter lacks beat and downbeat Embeddings, and it is trained for 150 epochs on the non-diffusion loss. It shows small improvements over the Beat This model on all metrics.

For the MDM, we observe substantial improvements in CMLt and AMLt metrics, which are particularly sensitive to erratic tempo changes. Strikingly, non-diffusion is outperformed even with a single inference step, indicating that the MDM training objective is beneficial beyond enabling iterative inference. We hypothesise that partially masked context reduces training ``confusion'' when the network is trained on similar pieces with conflicting metrical levels.

We also observe a positive impact of ensembling, which, aside from beat F1, improves all metrics and prevents metric degradation when using more inference steps. This is not a trivial result; averaging output probabilities could bring them closer to the binary decision boundary, thus increasing the likelihood of unwanted label flips. 
We may expect that more than 3 models would further improve performance (at the cost of inference time), but leave this study for future work.


\subsection{Comparison with current SOTA systems}

In Table~\ref{tab:soa}, we compare our ensemble performance with Gagneré et al.~\cite{gagnere2025}, and Ru et al.~\cite{ganghui2025hinge}, which currently have the highest performance in the literature. While direct comparison is not our objective, as these works focus on pre-training research orthogonal to our approach, we include these results to provide context for our metrics. Both papers use DBN post-processing in their best configuration, but \cite{gagnere2025} also reports metrics for a model that works without the DBN; their code and models are not public for a more in-depth evaluation or a statistical significance study. \cite{gagnere2025} uses a similar-sized model as ours, while \cite{ganghui2025hinge} fine-tunes a large audio model, MusicFM, which has more than 10x the parameters. We also include Beat This~\cite{foscarin2024beatthis} as our starting point, providing a clear indication of how much the techniques presented in this paper improve performance. We select 8 inference steps for our system, aiming to balance inference time and accuracy. We train our system 9 times with different seeds, aggregate them into 3 ensembles, compute 3 sets of scores, and report their means and standard deviations.

The comparison with systems without DBN post-processing shows that the MDM approach greatly improves performance, with consistent gains across all metrics, with up to $\approx 9 \%$ increases in downbeat CMLt and AMLt metrics, most sensitive to incoherent output behaviour.
Compared to DBN systems, we surpass the similarly sized system~\cite{gagnere2025}, and draw (in terms of the number of winning metrics) with the 10 times bigger model~\cite{ganghui2025hinge}.
It's worth recalling that, while GTZAN contains mostly pieces with 3 and 4 beats per bar, and a few time signature changes, the advantage of not using DBN post-processing is that we can expect our system to work on more diverse music with different time signatures, and time signature changes~\cite{foscarin2024beatthis}.

\subsection{Coherency measures}

We use two heuristics to evaluate the coherency of the output: we count consecutive downbeats (i.e., the number of downbeats whose previous beat position also contains a downbeat) and tempo doubling/halving changes (if an inter-beat interval is double or half of the preceding one, considering the same tolerance range used for the AMLt metric). These two factors do not make the output automatically wrong\footnote{Scriabin's Sonata 5 from the ASAP datasets contains a section with time signature 1/2; and tempo doubling/halving occurs, for example, in 4/4-7/8 tempo changes, and severe rallentando/accelerando.} but their occurrences are rare in GTZAN labels (0 consecutive downbeats and 0.033 doubling/halving).
We observe that both values decrease as the number of inference steps increases. For the 8-step inference, on average, per track, consecutive downbeat count drops from 0.25 to 0.02, and tempo doubling/halving occurrences pass from 0.75 to 0.119 (and decrease further to 0.06 for 20 steps). Manual evaluation of some pieces, such as the one from Figures~\ref{fig:problematic} and \ref{fig:inference}, shows that incoherences are completely fixed, with the outputs converging to a stable time signature and tempo.

\section{Conclusions}

We targeted the problem of incoherent NN outputs of beat trackers and presented a novel approach based on masked diffusion, which substantially reduces this problem and improves performance (notably on CMLt and AMLt scores), without requiring heavy post-processing based on simplistic musical assumptions. We achieve a new state-of-the-art performance across similarly sized systems, whether or not they use a DBN. These results were enabled by original modifications to standard masked diffusion techniques that addressed specific challenges of the beat-tracking task. 
The downsides, compared to traditional one-step approaches, are longer training (we double the number of epochs) and longer inference time (roughly $S$ times longer, even though we could optimise it by computing the frontend only once). Model ensembling also increases these costs linearly with the number of models.

We focused on a no-DBN approach; however, some applications require outputs that adhere to strict constraints. Since our method improves the quality of NN predictions, it could also enhance the performance of DBN-based systems. 
Beyond improved accuracy, our model's ability to accept partial outputs opens up practical use cases that one-step systems cannot support. A user can correct a set of beat positions and let the model complete the rest, which drastically speeds up the annotation of difficult pieces~\cite{sapinto2023challenging}.
Moreover, our model supports autoregressive and block-autoregressive inference without any training modification~\cite{nie2025llada}. We can achieve consistency on longer pieces, normally processed as independent snippets, by using the last part of a snippet output as a (always unmasked) conditioning for the next snippet prediction. Moreover, while we focused on obtaining a single coherent output from multiple valid interpretations, future work could explore techniques to control which interpretation is selected and allow switching between different metrical hierarchies. New evaluation protocols and new multi-annotation datasets are necessary to score such results correctly. Finally, our findings are general and can be applied to other beat trackers or even to other MIR tasks that may suffer from the multiple valid outputs problem, such as chord recognition and structure segmentation.

\section{Acknowledgments}
We would like to thank Jan Schlüter for the many productive discussions throughout this work, and for originally proposing iterative inference for beat tracking. We also owe a special thanks to Mathias Rose Bjare, who introduced us to the modern masked diffusion literature, sparking the very idea for this paper.
\bibliography{ISMIRtemplate}

%
%
%
%

\end{document}